\documentclass[prd,twocolumn,aps,letterpaper,amsmath,amssymb,preprintnumbers,showpacs,nolongbibliography,
floatfix,nofootinbib]{revtex4-2}

\usepackage{array}
\usepackage{graphicx}
\usepackage[export]{adjustbox}
\usepackage[caption=false]{subfig}
\usepackage[normalem]{ulem}
\allowdisplaybreaks 
\usepackage[hypertexnames=true]{hyperref}
\usepackage[capitalize]{cleveref}
\hypersetup{
    colorlinks=true,       
    linkcolor=blue,          
    citecolor=blue,        
    filecolor=blue,      
    urlcolor=blue           
  }

\usepackage{graphicx}
\usepackage{dcolumn}
\usepackage{bm}
\usepackage{xcolor}
\usepackage{slashed}

\usepackage{color}
\definecolor{BrickRed}{rgb}{0.8, 0.25, 0.33}
\definecolor{gray}{rgb}{0.6,0.6,0.6}
\definecolor{darkgreen}{rgb}{0.0, 0.545098, 0.0}
\definecolor{mypink1}{rgb}{0.858, 0.188, 0.478}

\begin{document}
\preprint{FERMILAB-PUB-23-458-CSAID-ND-T}

\title{Interfacing Electron and Neutrino Quasielastic Scattering Cross Sections with the Spectral Function in GENIE}

\author{Minerba Betancourt}
\author{Steven Gardiner}
\author{Noemi Rocco}
\author{Noah Steinberg}
\affiliation{Fermi National Accelerator Laboratory, P.O. Box 500, Batavia, IL 60410, USA}

\begin{abstract} Progress in neutrino-nucleus cross section models is being driven by the need for highly accurate predictions for the neutrino oscillation community. These sophisticated models are being developed within a microscopic description of the nucleus with the goal of encompassing all reaction modes relevant for the accelerator neutrino program. The disconnect between these microscopic models and the event generators that will be used in the next generation of experiments represents a critical obstacle that must be overcome in order to precisely measure the neutrino oscillation parameters. To this end we have developed a Fortran wrapper for lepton-nucleus quasielastic (QE) scattering within the GENIE event generator as a proof of principle, with the broader goal of creating an efficient pipeline for incorporating advanced theoretical models in event generators. As a demonstration of this interface, we have implemented the Spectral Function model into GENIE, offering a more complete description of the nuclear ground state, as well as the ability to provide quantifiable theoretical uncertainties. We validate this implementation and compare its predictions against data and against QE models already available in GENIE.
\end{abstract}

\maketitle

\section{Introduction}
The next generation of large accelerator-based neutrino oscillation experiments, namely DUNE and Hyper-K, will require an evolution in our understanding and modeling of neutrino-nucleus interactions in order to meet their design goals~\cite{DUNE:2020ypp,Hyper-Kamiokande:2018ofw}. These experiments aim to not only measure the standard neutrino oscillation parameters, but also to challenge the three neutrino paradigm and search for other physics beyond the Standard Model~\cite{DUNE:2020fgq,Brdar:2022vum}. This requires accurate predictions for all SM (and BSM) processes as well as a quantification of the associated systematic errors involved. These experiments rely on neutrino event generators for the above, which makes the accuracy of such generators of paramount importance. Fortunately, modern neutrino event generators have a plethora of new lepton-nucleus scattering data to benchmark against with higher precision~\cite{MINERvA:2019gsf,T2K:2020jav,CLAS:2021neh}, in new exclusive channels~\cite{MINERvA:2019ope,MINERvA:2021wjs,T2K:2021naz,MicroBooNE:2023dwo,MicroBooNE:2022emb}, with highly differential data~\cite{MINERvA:2022mnw}, and on new targets~\cite{MicroBooNE:2023jyj,MINERvA:2023kuz}. These new results have without a doubt shown that the empirical models used in many event generators cannot simultaneously describe the data across the landscape of experiments.

A common practice among generators is to stitch together disparate models, each describing a different reaction mechanism -- quasielastic, two-particle two-hole (2p2h), resonance production, deep inelastic scattering. These are woven together to cover the phase space probed by neutrino experiments. The lack of a unified framework for each of the components leads to large ad hoc tunes being applied, interaction by interaction, to reach agreement with the data~\cite{MINERvA:2015ydy,NOvA:2020rbg,GENIE:2021zuu,MicroBooNE:2021ccs}. These tunes tend to give inconsistent results across nuclear targets, and even across experiments using the same nuclear target. Additionally, such empirical treatments provide no way to rigorously assess the theoretical uncertainty associated with the underlying physics, obscuring the final systematic errors obtained on the sought after oscillation parameters.  

Reaching the $\mathcal{O}(1)\%$ precision in the neutrino cross section predictions needed for neutrino oscillation analyses will require basing our models in first principles nuclear theory, a consistency in the treatment of the different reaction mechanisms relevant to describe experimental data, and the implementation of such models in our event generators to estimate signal and background predictions~\cite{DUNE:2015lol}. In this article, we will describe a new interface developed for the GENIE event generator~\cite{GENIE:2021npt,Andreopoulos2010} which enables an efficient implementation of the Spectral Function model for the description of the quasi-elastic region. This interface can be easily adapted to other accommodate other nuclear models.

The Spectral Function and extended factorization scheme allow for a unified framework able to describe the different  reaction mechanisms into the same model while providing an accurate description of nuclear dynamics. Furthermore, it allows to consistently estimate the theoretical error of the calculations, preliminary studies in this direction have been carried out in 
Ref.~\cite{Rocco:2018mwt,Simons:2022ltq}. Section~\ref{sec:theory_interface} discusses the motivation of and implementation of a theory interface, while Sec.~\ref{sec:factorization} and~\ref{sec:spectral_func} give details on the factorization scheme and Spectral Functions used in the model. Finally in Sec.~\ref{sec:validation} and~\ref{sec:exclusive} we validate and test the implementation against inclusive and exclusive electron and nucleus scattering data.

\section{Theory Interface}\label{sec:theory_interface}
As the number and sophistication of lepton-nucleus interaction models grow, one of the most time consuming bottlenecks is the implementation of these models into event generators. Currently this must be done by a specialist, with specific knowledge of a particular event generator. Models are typically added one at a time, often requiring both translation between programming languages and adaptation to existing software infrastructure. An example of this is the SuSAv2 implementation in GENIE in which the theoretical model is designed only for inclusive interactions, but the event generator must be able to deliver fully exclusive predictions~\cite{Dolan:2019bxf}.

The need for a less labor-intensive pipeline for theorists to contribute models to event generators has motivated development of simple interfaces for integrating external calculations~\cite{Barrow:2020gzb}.
In the GENIE neutrino event generator a first step in this direction was taken through the creation of a hadron tensor table framework~\cite{GENIE:2021npt}. In this framework pre-computed tables of hadronic response tensor elements, defined on a two-dimensional grid in energy and momentum transfer, are provided to GENIE for sampling of the final-state lepton kinematics. The hadron tensor can be contracted with a generic leptonic tensor to compute either charged lepton or neutrino scattering cross sections. The tensor table framework has been adopted for the inclusion of the CRPA QE model, SuSAv2 QE+2p2h model, and the Valencia 2p2h model~\cite{Gonzalez-Rosa:2022ltp, Gran:2013kda, Nieves:2011pp,Dolan:2021rdd}. 

While the tensor table strategy allows for a speedy implementation of these models into GENIE, the framework has several drawbacks. The current GENIE format for tensor tables is inclusive, meaning that the outgoing nucleon kinematics must be sampled separately from those of the final-state lepton. This has the potential to lead to large disagreements in nucleon momentum and angle distributions~\cite{Nikolakopoulos:2023pdw}. Additionally, there are questions of consistency between the underlying nuclear ground state used to generate the tensor tables and the ground state used in GENIE to select target nucleons. Finally, there is no ability to manipulate the underlying theory parameters involved in the calculation of the hadron tensor elements themselves. This ability can be useful for studying systematic uncertainties, which must otherwise be estimated by less well-motivated methods. 

As a first step towards a more flexible interface which addresses these challenges, we have removed the barrier between theorists' original codes and GENIE by creating a Fortran wrapper to directly interface these codes with the GENIE event generator. The choice to create a Fortran wrapper as opposed to any other programming language was based on a survey of many theorists in the neutrino-nucleus scattering community in which a majority of theorists had implementations of their models written in Fortran~\cite{ModelSurvery}. The first wrapper developed is specifically for predictions of QE scattering within the Impulse Approximation (IA). In this scheme, described further in Secs.~\ref{sec:factorization} and~\ref{sec:spectral_func}, lepton-nucleus scattering is factorized into the incoherent sum of collisions with individual nucleons. The nuclear ground state is described by a probability density known as the Spectral Function (SF) which specifies the energy and momentum distributions of bound nucleons. Realistic Spectral Functions include both short- and long-range correlations between constituent nucleons. Given an input Spectral Function, our wrapper allows for a calculation of the hadronic response tensor from an external theory code written in Fortran. This capability can then be used by GENIE to produce events and compute differential cross sections. In the following sections we will give more detail about the factorization scheme used; contrast the Spectral Function against other, more simple nucleon momentum distributions; and validate and compare the model predictions against charged lepton and neutrino scattering data.

\section{Factorization of Electron and Neutrino quasielastic Scattering}
\label{sec:factorization}
We report the expression of the fully exclusive lepton-nucleus differential cross section yielding single-nucleon emission. Within
the IA, which is expected to hold for momentum transfers $|\textbf{q}| > 400\,\rm{MeV}$, this can be written in the form
\begin{equation}\label{eq:xsec}
\begin{aligned}
d\sigma &= \sum_{\tau = n,p}\frac{\mathcal{N}_{\tau}\mathcal{C}}{ 32\pi^{2}E_{\textbf{p}}E_{\textbf{p}'}E_{\textbf{k}'}E_{\textbf{k}} }P_{\tau}(\textbf{p},E)\times \\
&L_{\mu\nu}\tilde{A}_{\tau}^{\mu\nu}\delta( E_{\textbf{k}} + E_{N_{i}} - E_{\textbf{k}'} - E_{\textbf{p}'})\,d^{3}\textbf{p}\,dE\,d^{3}\textbf{k}'\, .
\end{aligned}
\end{equation}
In Eq.~\ref{eq:xsec} $k$ $(k')$ and $p$ $(p')$ denote the four-momenta of the initial (final) lepton and initial (final) struck nucleon, respectively, and $E_{\textbf{p}}$ is the on-shell energy of a particle with 3-momentum $\textbf{p}$. 
The leptonic tensor is completely determined by the lepton kinematics and is given separately for charged leptons and neutrinos as
\begin{equation}
  L_{\mu\nu} =
    \begin{cases}
      \rm{CC, NC} & 8(k_{\mu}k'_{\nu} + k'_{\mu}k_{\nu} - k\cdot k'g_{\mu\nu} \pm i\epsilon_{\mu\nu\rho\sigma}k^{\rho}k'^{\sigma})\\
      \rm{EM} & 2(k_{\mu}k'_{\nu} + k'_{\mu}k_{\nu} +[m_{\textbf{k}}^{2} - k\cdot k'g_{\mu\nu}])\, .\\
    \end{cases}       
\end{equation}
The upper sign (+) should be taken for neutrinos and the lower (-) for anti-neutrinos. We use the symbol $m_{\textbf{k}}$ to represent the mass of the particle with 3-momentum $\textbf{k}$. The coupling factor $\mathcal{C}$ depends on the probe and is given by
\begin{equation}
    \mathcal{C} = 
    \begin{cases}
        \rm{CC} & G_{F}^{2}|\rm{V}_{\rm{ud}}|^{2}\\
        \rm{NC} & G_{F}^{2}\\
        \rm{EM} & 2e^{4}/Q^{4}\, ,\\
    \end{cases}
\end{equation}
where $Q^{2} = -q^{2} > 0$.
\vskip 0.12in
$\tilde{A}_{\tau}^{\mu\nu}$ is the nucleon-level response tensor for a bound nucleon with isospin $\tau$. In the IA, $\tilde{A}^{\mu\nu}$ is just the free nucleon response tensor but with the energy transfer $\omega$ modified to account for the energy that must be given to the residual nucleus to free the bound nucleon,
\begin{equation}\label{eq:nucleonresponse}
    \tilde{A}^{\mu\nu} = \langle p'|j^{\dagger\mu}_{1b}(\tilde{q})|p\rangle\langle p|j^{\nu}_{1b}(\tilde{q})|p'\rangle.
\end{equation}
In the above we have assumed that the nuclear current operator is made up of only one-body currents, i.e.,
\begin{equation}\label{eq:nuc_current}
    J^{\mu}_{\rm{nuclear}} = \sum_{i}j^{\mu}_{1b}\, .
\end{equation}
The single nucleon Spectral Function $P_{\tau}(\textbf{p},E)$ describes the distribution of momentum and removal energy for bound nucleons of isospin $\tau$. Asymmetric nuclei like $^{40}$Ar necessarily have different Spectral Functions for protons and neutrons, so it is important that Eq.~\ref{eq:xsec} allow for this. For the case of symmetric nuclei we can ignore isospin breaking effects and easily set $P_{p}(\textbf{p},E) = P_{n}(\textbf{p},E)$. The binding energy of each nucleon is given by $\epsilon_{B} = M_{f} + m_{p} - M_{i}$ where $M_{f(i)}$ is the mass of the final (initial) nucleus. In Eq.~\ref{eq:nucleonresponse} we follow the DeForest prescription of using free nucleon spinors and form factors, evaluated using on shell nucleon four-momenta but a modified four momentum transfer~\cite{DeForest:1983ahx}
\begin{equation}
    \tilde{q} = p' - (E_{\textbf{p}}, \textbf{p}) = q - (\epsilon_{B},\textbf{0}) = (\tilde{\omega},\textbf{q})\, .
\end{equation}
The nucleon current operator is given by
\begin{equation}\label{eq:current}
\begin{aligned}
j^{\mu}_{1b} &= \gamma^{\mu}F_{1}^{V}(\tilde{Q}^{2}) + i\sigma^{\mu\nu}\frac{\tilde{q}_{\nu}}{2M}F_{2}^{V}(\tilde{Q}^{2}) \\
&+ \gamma^{\mu}\gamma^{5}F_{A}(\tilde{Q}^{2}) + \frac{\tilde{q}^{\mu}}{M}\gamma^{5}F_{P}(\tilde{Q}^{2}).
\end{aligned}
\end{equation}
Finally, the form factors used in Eq.~\ref{eq:current} in the case of charged lepton scattering are related to those used in neutrino scattering by the Conserved Vector Current (CVC) hypothesis. This relationship allows for vector form factors derived from precision electron scattering experiments to be readily implemented in neutrino-nucleus cross section predictions. Several parameterizations of the Dirac and Pauli form factors $F_{1,2}^{p,n}$ exist in GENIE which can be configured by the user. For the axial form factor we consider the dipole model with $M_{A} = 1.0$ GeV~\cite{Bodek:2007ym}, but the z-expansion parameterization extracted from neutrino-Deuterium scattering~\cite{Meyer:2016oeg} as well as from Lattice QCD~\cite{RQCD:2019jai,Park:2021ypf,Djukanovic:2022wru} also exists in GENIE. Obviously for charged lepton scattering we set $F_{A} = F_{P} = 0$.
\vskip 0.12in
This model simultaneously describes both charged lepton and neutrino-nucleus scattering.
Comparisons against inclusive and semi-exclusive electron scattering data have already highlighted several modeling deficiencies in the current generation of neutrino event generators~\cite{electronsforneutrinos:2020tbf,CLAS:2021neh}. 
\section{Spectral Function}\label{sec:spectral_func}
The Spectral Function of a nucleon with isospin $\tau\in\{p,n\}$ and momentum ${\bf k}$ can be written as
\begin{align}
    P_{\tau}(\mathbf{k},E)&=\sum_n |\langle 0| [|k\rangle\, |\Psi_n^{A-1}\rangle]|^2 \delta(E+E_0-E_{n}^{A-1})\nonumber\\
    &=P_{\rm{MF}}(\mathbf{k},E) + P_{\rm{corr}}(\mathbf{k},E)\, ,
\label{pke:hole}
\end{align}
where $|k\rangle$ is the single-nucleon, plane-wave state, $|0\rangle$ is the ground state of the Hamiltonian with energy $E_0$, while $|\Psi_n^{A-1}\rangle$ and $E_{n}^{A-1}$ are the energy eigenstates and eigenvalues of the remnant nucleus with $(A-1)$ particles. The Spectral Function in Eq.~\ref{pke:hole} is a sum of a mean field (MF) and a correlation (corr) term with distinct energy dependence. Both exclusive and inclusive electron scattering experiments have shown that the correlation piece dominates for momenta above $k_{f}$, is essentially universal, and comprises approximately 20\% of the single nucleon strength~\cite{CLAS:2005ola,CLAS:2022odn,Weiss:2020bkp,Hen:2014nza,JeffersonLabHallA:2007lly,CLAS:2020rue}. The momentum distribution of the initial nucleon is obtained by integrating the Spectral Function over the removal energy  
\begin{equation}
n_{\tau}(k) = \int dE\, P_{\tau}(\mathbf{k},E)\,.
\end{equation}
\vskip 0.12in
Nuclear models currently included in GENIE are based on either the Relativistic Fermi Gas (RFG) or the Local Fermi Gas (LFG), the latter of which uses a density-dependent Fermi momentum. Ad-hoc modifications of the models include fixed high momentum tails stitched onto the original RFG momentum distributions~\cite{Bodek:1981wr} or (in the LFG) a shift in strength from $k<k_{f}$ to $k>k_{f}$~\cite{GENIE:2021npt}, mimicking a correlation tail. In either case this leads to an incorrect relationship between the nucleon momentum and removal energy. Spectral Functions for finite nuclei have been derived from experiment and different theoretical approaches (QMC, LDA, SCGF, CBF)~\cite{JeffersonLabHallA:2022cit,CLAS:2022odn,Benhar:1994hw,Benhar:1989aw,Soma:2012zd}. In this work we utilize the $^{12}\rm{C}$ and $^{16}$O Spectral Functions obtained within the Correlated Basis Function (CBF) approach, where the MF piece has been fit to $(e,e'p)$ scattering data and the correlation contribution is computed using the Local Density Approximation (LDA)~\cite{Benhar:1994hw}. We also assume that the SF for protons and neutrons are the same, and we ignore any isospin breaking effects. We note that the availability of several Spectral Functions from different underlying nuclear models is an advantage as it presents an opportunity to quantify related theoretical uncertainties.

\begin{figure}[h]
    \centering
    \includegraphics[width=\linewidth]{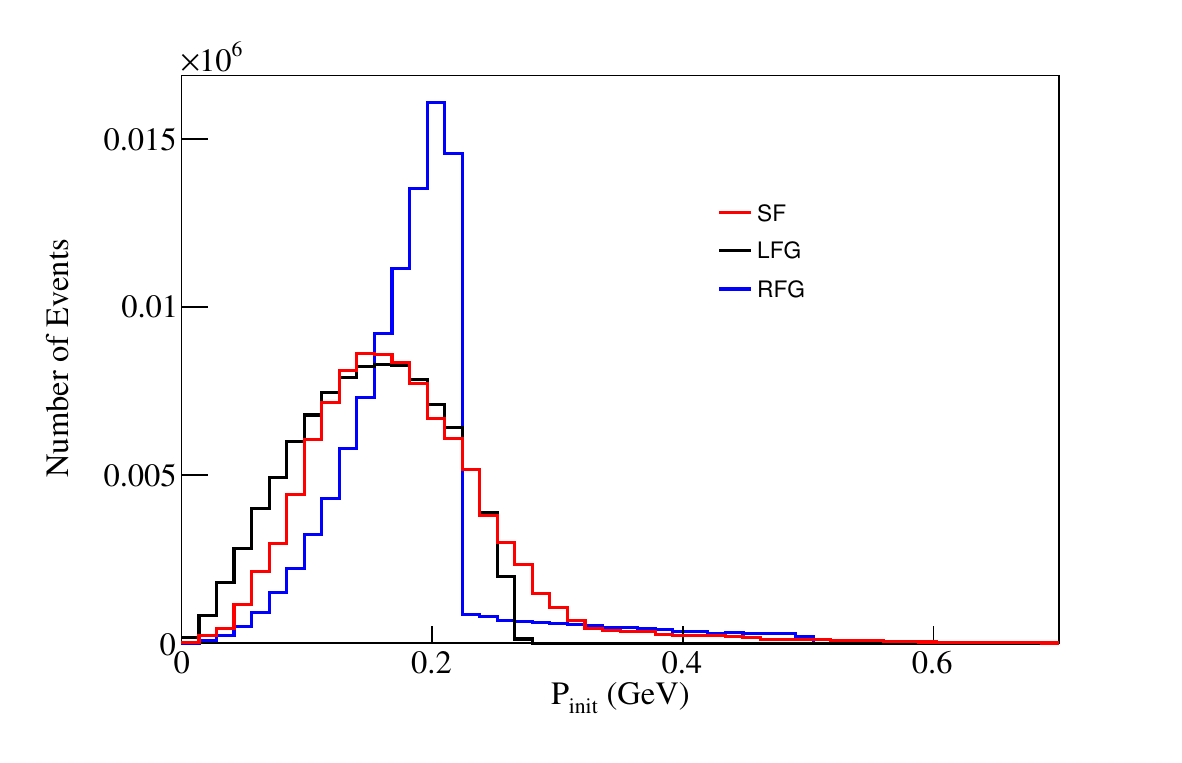}
    \caption{Initial nucleon momentum distributions for $^{12}\rm{C}$ for models using the Relativistic Fermi Gas (RFG: Blue), Local Fermi Gas (LFG: Black), and Spectral Function (SF: Red). Momentum distributions have been obtained from 100,000 simulated electron $^{12}$C scattering events at $E_{\rm{beam}} = 1\,\rm{GeV}$ in GENIE.}
    \label{fig:momentum_dist}
\end{figure}

Figure~\ref{fig:momentum_dist} displays the initial-state nucleon momentum distribution for true QE events on $^{12}\rm{C}$ produced using GENIE and the RFG, LFG, and Spectral Function representations of the target nucleus.
It is clearly visible that the normalization of the RFG in the mean field (low momentum) region is much larger than the LFG and SF. The LFG completely lacks the correlation tail which is put in by hand in the RFG but exists $\textit{a priori}$ in the SF. Measurements from MINERvA and T2K of single transverse kinematic imbalance observables have shown that the largest disagreement between models exists in this SRC dominated region between $200 < p_{n} < 700$~MeV. The SF initial state agrees better with the data in this region than nuclear models based on the RFG and LFG~\cite{Dolan:2018zye,T2K:2018rnz}.

\section{GENIE Implementation}
\begin{figure}
    \centering
    \includegraphics[width=\linewidth]{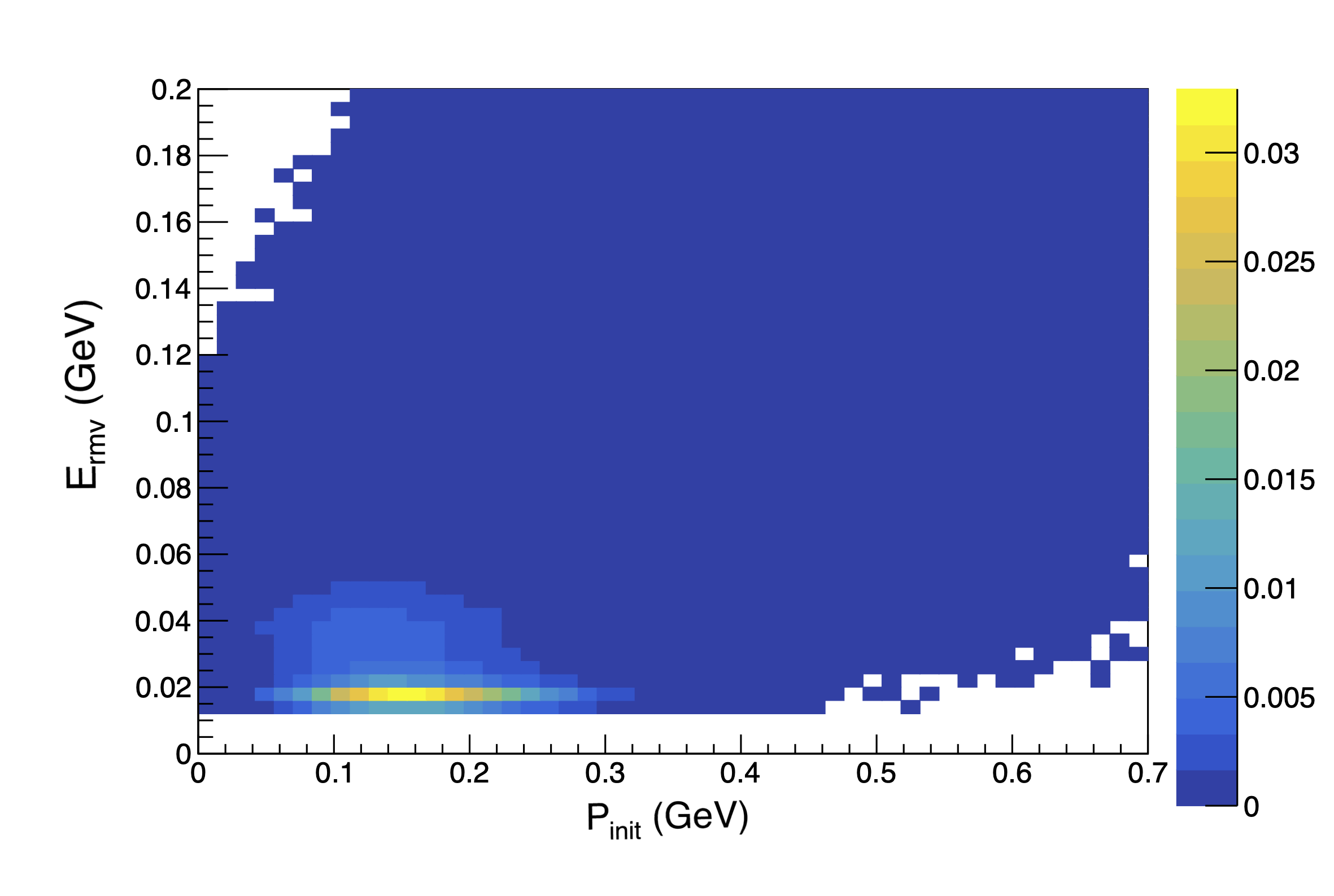}
    \caption{Two dimensional probability mass distribution of initial nucleon momentum and removal energy for the $^{12}\rm{C}$ SF implemented in GENIE. S and P shells are visible at low momentum and removal energy.}
    \label{fig:2Dhist}
\end{figure}

The first step of the GENIE implementation involves some minor code adjustments to allow use of a precomputed SF provided in the form of a data file. Each SF data file contains a table of $|\textbf{p}|,E,P(|\textbf{p}|,E)$ triples arranged on a regular grid. The SF is normalized so that 
\begin{equation}\label{eq:SF}
\begin{aligned}
    &\int P(\textbf{p},E)d^{3}\,\textbf{p}\,dE\\
    &\approx 4\pi\Delta|\textbf{p}|\Delta E\sum_{ij}|\textbf{p}_{i}|^{2}P(|\textbf{p}_{i}|,E_{j})\\
    &= \sum_{ij}P_{\rm{bin\,ij}} = 1,
\end{aligned}
\end{equation}
where $|\textbf{p}_{i}|$ and $E_{j}$ are evaluated at the midpoint of each bin on the grid.
The values of $|\textbf{p}|$ and $E$ are sampled for an initial nucleon using a two-dimensional histogram like the one shown in Fig.~\ref{fig:2Dhist}. The bins of this histogram have been filled with the same probability mass value $P_{\rm{bin\,ij}}$ from Eq.~\ref{eq:SF}. To approximate the SF, a 2D bin is sampled according to the probability mass distribution, and then specific values of $|\textbf{p}|$ and E are chosen uniformly within its boundaries. Finally, a direction for the initial nucleon  is chosen isotropically.

New code was also added to GENIE (in the form of a C++ class called $\texttt{UnifiedQELPXSec}$) to compute the quasielastic differential cross section according to the expression from Eq.~\ref{eq:xsec}. The new code takes advantage of the flexibility of the formalism in Sec.~\ref{sec:factorization} to simultaneously describe electron and neutrino scattering. Based on the projectile of interest, GENIE sets up any necessary constants, form factors, or other calculation ingredients from GENIE internals, minimizing the need for code duplication. Utilizing the same model and code for charged lepton and neutrino scattering allows for parameter constraints obtained from charged lepton scattering experiments to be consistently and immediately applied to neutrino scattering (as well as vice versa). Our implementation utilizes the wrapper described in Sec.~\ref{sec:theory_interface} to compute the nucleon-level response tensor of Eq.~\ref{eq:nucleonresponse} using an external Fortran code. The results are then fed back to GENIE to compute the differential cross section.

In order to remove the energy-conserving delta function of Eq.~\ref{eq:xsec}, we utilize a change of variables by working within the center of momentum (CM) frame of the initial lepton and the struck nucleon. In this reference frame, a formal replacement can be made
\begin{equation}\label{eq:jacobian}
\begin{aligned}
    &\delta( E_{\textbf{k}} + E_{N_{i}} - E_{\textbf{k}'} - E_{\textbf{p}'})\,d^{3}\textbf{k}'\rightarrow\nonumber\\
    &\frac{\sqrt{1 + (\gamma^{2} -1)(1-\cos^{2}\theta_{0})}}{|\mathbf{v}_{\mathbf{k}'} - \mathbf{v}_{\mathbf{p}'}|}|\mathbf{k}'_{0}|^{2}\,d\phi_{0}\,d\cos\theta_{0}\, ,
\end{aligned}
\end{equation}
where $\mathbf{k}'_{0}$ is the final lepton 3-momentum in the CM frame, $\gamma$ is the Lorentz factor for the boost between lab and CM frames, and $\mathbf{v}_{\mathbf{k}'}$ ($\mathbf{v}_{\mathbf{p}'}$) is the lab-frame velocity of the final lepton (final nucleon). The CM frame final lepton scattering angles $\theta_{0}$ and $\phi_{0}$ are measured between $\mathbf{k}'_{0}$ and $\mathbf{v}$, the velocity of the CM frame as measured in the lab frame. This choice of variables is convenient for Monte Carlo (MC) sampling, and is also done for existing QE simulations in recent releases of GENIE.

By using the Spectral Function as a normalized probability density, we can integrate over the 4D phase space of the initial nucleon using MC methods. The differential cross section can be computed as
\begin{equation}
    \begin{aligned}
        \frac{d\sigma}{d\cos\theta_{0}d\phi_{0}} &= \int P(\mathbf{p},E)F(\mathbf{p},E)dEd^{3}\mathbf{p}\nonumber\\
        & = \langle F(\mathbf{p},E)\rangle \approx \frac{1}{N}\sum_{k=1}^{N}F(\mathbf{p}_{k},E_{k})\, ,
    \end{aligned}
\end{equation}
where $F(\mathbf{p}_{k},E_{k})$ is basically the cross section of Eq.~\ref{eq:xsec} with the Spectral Function factored out. Nucleon variables are drawn for each trial from the Spectral Function, and the lepton angles are easily integrated over.

While the above constitutes a novel implementation of the Spectral Function into GENIE, we must mention past work on another numerical implementation called GENIE + $\nu T$~\cite{Jen:2014aja}. This work focused on inclusive observables and studied the corresponding shift in extracted oscillation parameters when the SF is used as the base model. While the origin of the physics model in the GENIE + $\nu T$ implementation is the same as in the present work, several differences must be noted. First, the kinematic sampling is done differently. In Ref.~\cite{Jen:2014aja}, values of $Q^{2}$ are generated for sampling the lepton kinematics, as was typical in GENIE releases before major version 3; our implementation generates $(\cos\theta_{0},\phi_{0})$ pairs which fully retain correlations with the outgoing nucleon. This enables our implementation to deliver exclusive cross section predictions needed for analyzing the data of current and future oscillation experiments using liquid argon time projection chambers. The goals of our implementation are also different. First and foremost, the present work serves as a test case for our Fortran wrapper and a verification that the implementation is done correctly. Our SF implementation is also part of a larger effort to improve lepton-nucleus scattering models in event generators, with a hope to develop a consistent scheme which encompasses all reaction mechanisms. One alternative avenue is the development of the ACHILLES event generator, also based on the SF model, which aims to root each portion of the event-generation pipeline in microscopic nuclear theory~\cite{Isaacson:2021xty,Isaacson:2022cwh}.  

\section{Validation}\label{sec:validation}
\begin{figure}[h]    
    \centering
    \includegraphics[width=\linewidth]{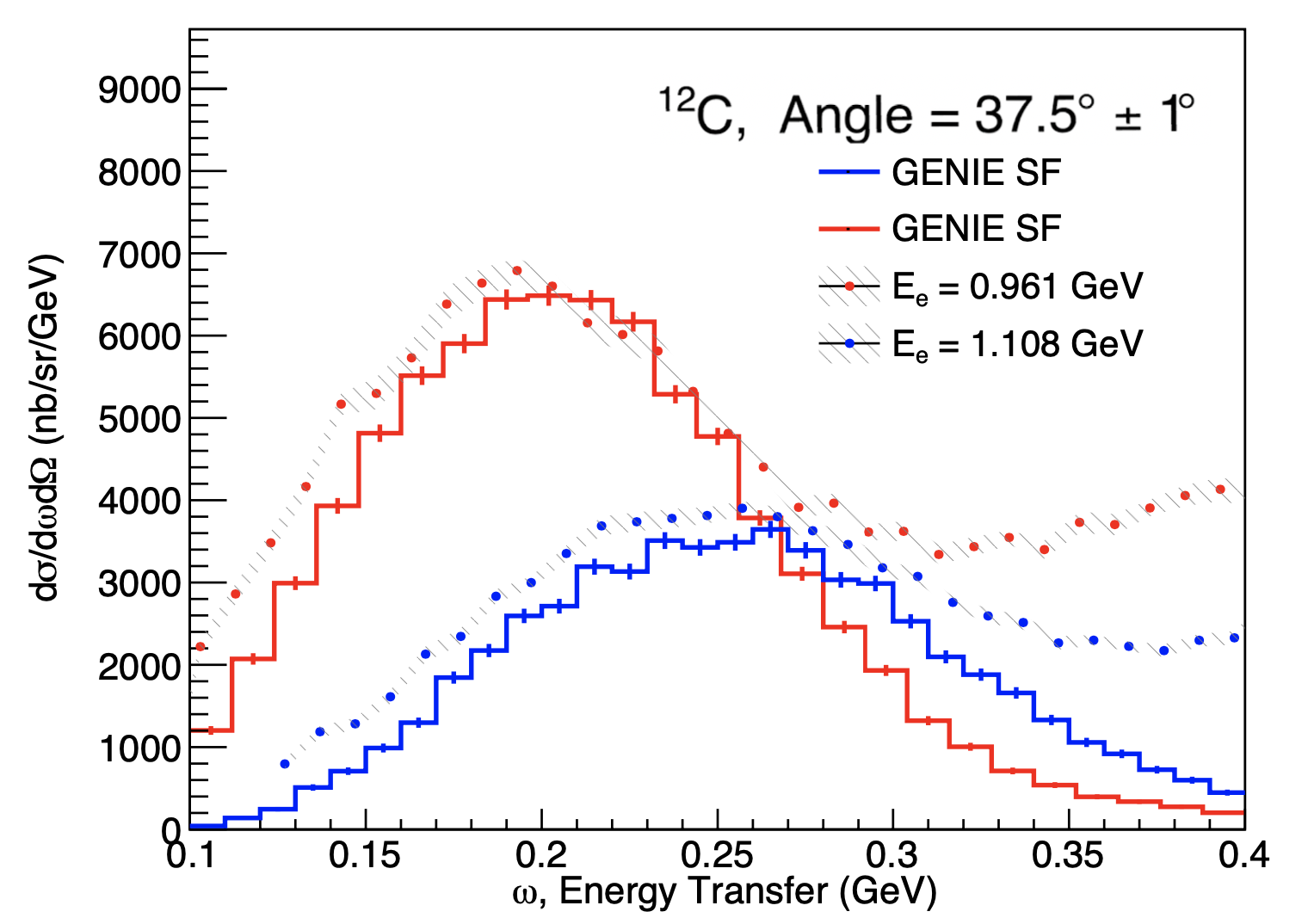}
    \caption{Inclusive double differential cross sections vs. energy transfer from $e$-$^{12}\rm{C}$ scattering at $\theta_{e'} = 37.5^\circ$ for beam energies of $0.961$ (red) and $1.108$ (blue) GeV. Data points are shown as points in the same colors with shaded bands showing statistical plus systematic errors }
    \label{fig:e_inclusive}
\end{figure}

As a validation of our implementation, we first compare our GENIE SF results against inclusive electron scattering data and standalone calculations (i.e., outside of any event generator) using the same Spectral Function and form factors for inclusive neutrino scattering. In Fig.~\ref{fig:e_inclusive} we show predictions using the GENIE SF model against inclusive electron scattering data on $^{12}\rm{C}$ for beam energies of $0.961$ and $1.108$ GeV, both taken at an electron scattering angle of $37.5^{\circ}$~\cite{Sealock:1989nx}.
We see here that the peak locations and widths are well described by the SF model, though final state interactions will slightly shift the peaks towards lower energy transfer through interference effects~\cite{Benhar:2013dq}. The GENIE SF model slightly under-predicts the height of the peaks, but this is to be expected. The inclusion of two-body currents in Eq.~\ref{eq:nuc_current} leading to multi-nucleon knockout increases the predicted cross section especially at energy transfers beyond the QE peak and before resonance production. Furthermore the interference between one- and two-body currents leading to single nucleon knockout is known to increase the cross section at the QE peak~\cite{Rocco:2015cil,Benhar:2013bba,Benhar:2015ula}. Given the missing interaction mechanisms just mentioned, the satisfactory agreement between the GENIE SF predictions and the inclusive data is a useful validation of the implementation. Below in Fig.~\ref{fig:numu_inclusive} we also show inclusive double differential muon neutrino cross sections at $E_{\nu} = 1\,\rm{GeV}$ at fixed muon scattering angles of $20^{\circ},30^{\circ}$, and $40^{\circ}$. Predictions from the GENIE SF match the standalone calculations (labeled ``Rocco SF'' in the figure), again validating the implementation.
\begin{figure}[h]    
    \centering
    \includegraphics[width=\linewidth]{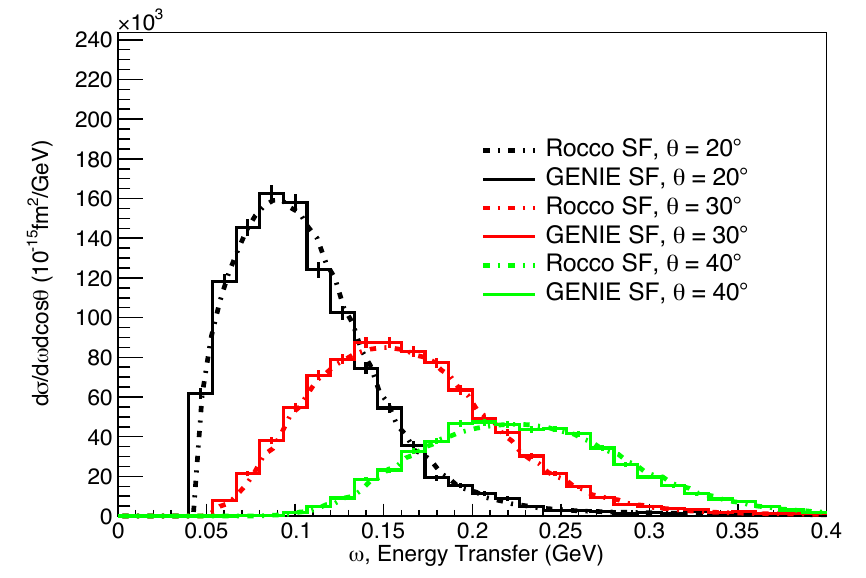}
    \caption{Inclusive double differential cross sections vs. energy transfer from $\nu_{\mu}$-$^{12}\rm{C}$ QE scattering at 1 GeV and several muon scattering angles: 20, 30, and 40 degrees. Solid lines are the GENIE SF implementation and the dashed lines are predictions from the SF model of Noemi Rocco.}
    \label{fig:numu_inclusive}
\end{figure}
\section{Exclusive Cross Section Predictions}\label{sec:exclusive}
\begin{figure}[h]
    \centering
    \includegraphics[width=\linewidth]{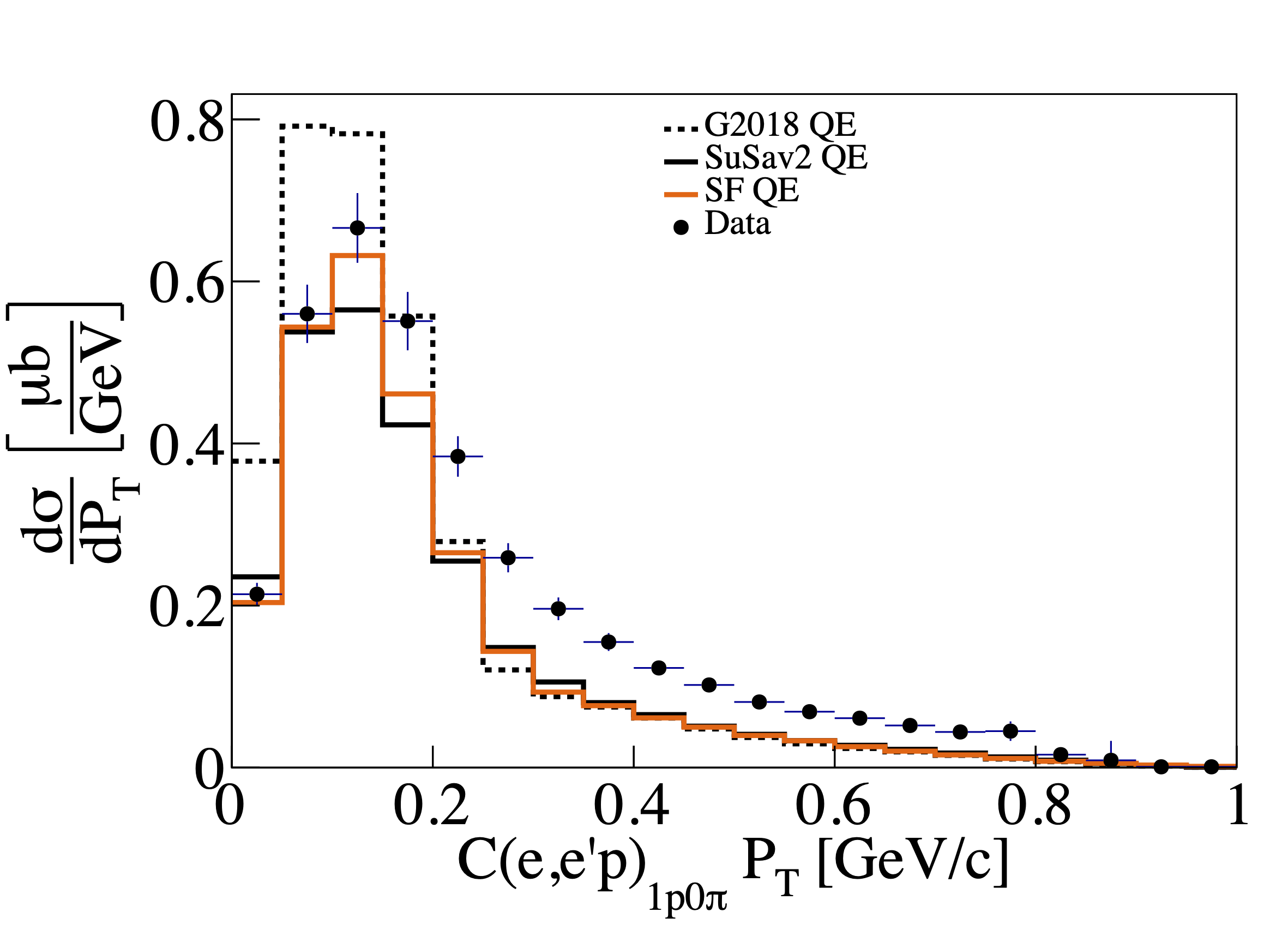}
    \caption{Differential cross section in $p_{T}$ from (e,e'p$)_{0\pi}$ events for $1.159$ GeV e- $^{12}\rm{C}$ GeV scattering. Simulation predictions from three different GENIE models where only true QE events are shown.}
    \label{fig:e4v}
\end{figure}

\begin{figure*}[t]
    \centering
    \includegraphics[width=\linewidth]{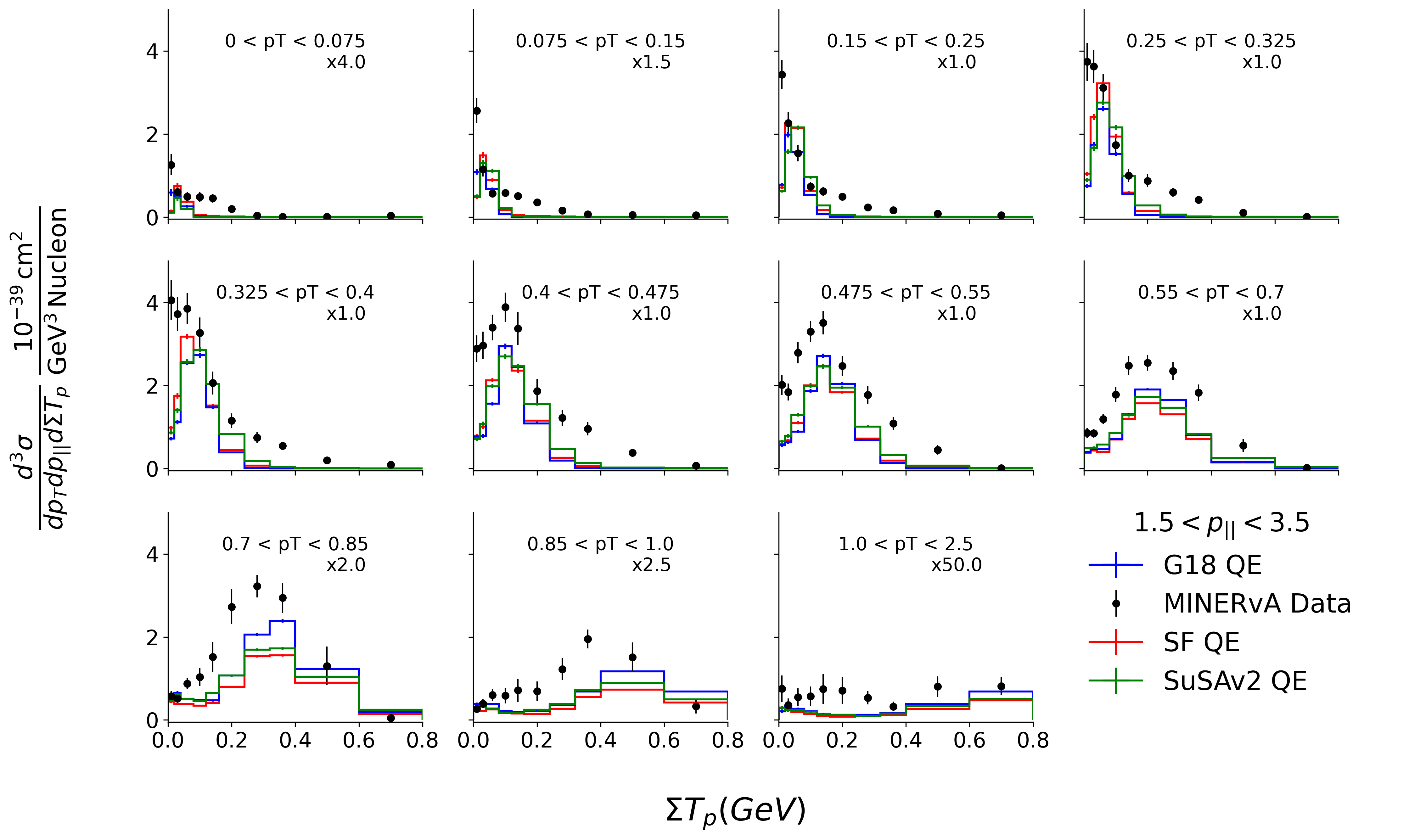}
    \caption{MINERvA CCQE-like triple differential cross section (black points) compared to SF (red), SuSAv2 (green), G18 (blue). Plot above is for $1.5 < p_{||}/\rm{GeV} < 3.5$, in bins of $p_{T}$ against $\sum T_{p}$ which is the sum of all observed protons' kinetic energy.}
    \label{fig:triple}
\end{figure*}

As discussed earlier, exclusive cross sections are a more powerful discriminator between different neutrino-nucleus cross section models. To this end we compare the GENIE SF implementation to both the SuSAv2 QE and G2018 QE models implemented in GENIE. We include only the quasielastic components of each model for consistency in the comparisons. We begin with exclusive electron scattering measurements from e4$\nu$, where the $(e,e'p)_{0\pi}$ topology has been measured on a variety of targets and across multiple beam energies~\cite{CLAS:2021neh}. We focus on transverse kinematic imbalance variables (TKI) which are sensitive to different reaction mechanisms and are independent of incident lepton energy~\cite{MINERvA:2018hba,T2K:2018rnz,Lu:2015tcr}. The differential cross section in transverse momentum defined as
\begin{equation}
    \rm{\mathbf{P}}_{\rm{T}} =  \rm{\mathbf{P}}^{e'}_{\rm{T}} + \rm{\mathbf{P}}^{p}_{\rm{T}}\, ,
\end{equation}
for $1.161$ GeV electrons on $^{12}$C compared to predictions from G2018, SuSAv2, and the SF are shown in Fig.~\ref{fig:e4v}. quasielastic scattering has been shown to be the dominant component at low $\rm{P}_{\rm{T}}$ where Fermi-motion dictates the normalization and width of the cross section. Inelastic contributions, NN correlations, and significant intra-nuclear re-scattering or re-absorption of the outgoing hadronic system (FSI) contribute as a broad tail to higher values of $\rm{P}_{\rm{T}}$ above the Fermi momentum.
Figure~\ref{fig:e4v} shows that G18 model significantly over predicts the normalization at low $\rm{P}_{\rm{T}}$. The SF model shows an excellent agreement with the data at low $\rm{P}_{\rm{T}}$ where it should be remarked that as the simulations include only the QE interaction, the predictions should $\textit{always}$ undershoot the data. The lowest $\rm{P}_{\rm{T}}$ bin shows a mild over prediction from the SuSAv2 model, but otherwise SuSAv2 describes the data well. The cross section serves as a proxy for initial nucleon momentum, as can be seen by the similarities between the shape and normalization of the cross sections in Fig.~\ref{fig:e4v} compared to the momentum distributions of Fig~\ref{fig:momentum_dist}.
\vskip 0.12in
Moving from electron scattering to neutrinos we next examine CCQE-like (also known as CC$0\pi$) scattering from the MINERvA experiment. The ability of a model to simultaneously describe electron and neutrino scattering is crucial to leveraging the extremely high precision charged lepton data available. To this end we examine first data from the Low Energy (LE) period of MINERvA, with an average neutrino energy $\langle E_{\nu} \rangle = 3\,\rm{GeV}$. We focus on another derived TKI variable, $p_{n}$ which is an estimator for the initial neutron momentum under the CCQE hypothesis~\cite{MINERvA:2018hba}. Below in Fig.~\ref{fig:pn} we show the measured $p_{n}$ distribution at MINERvA against QE predictions from SuSAv2 and the SF models. 
\begin{figure}[h]
    \centering
    \includegraphics[width=\linewidth]{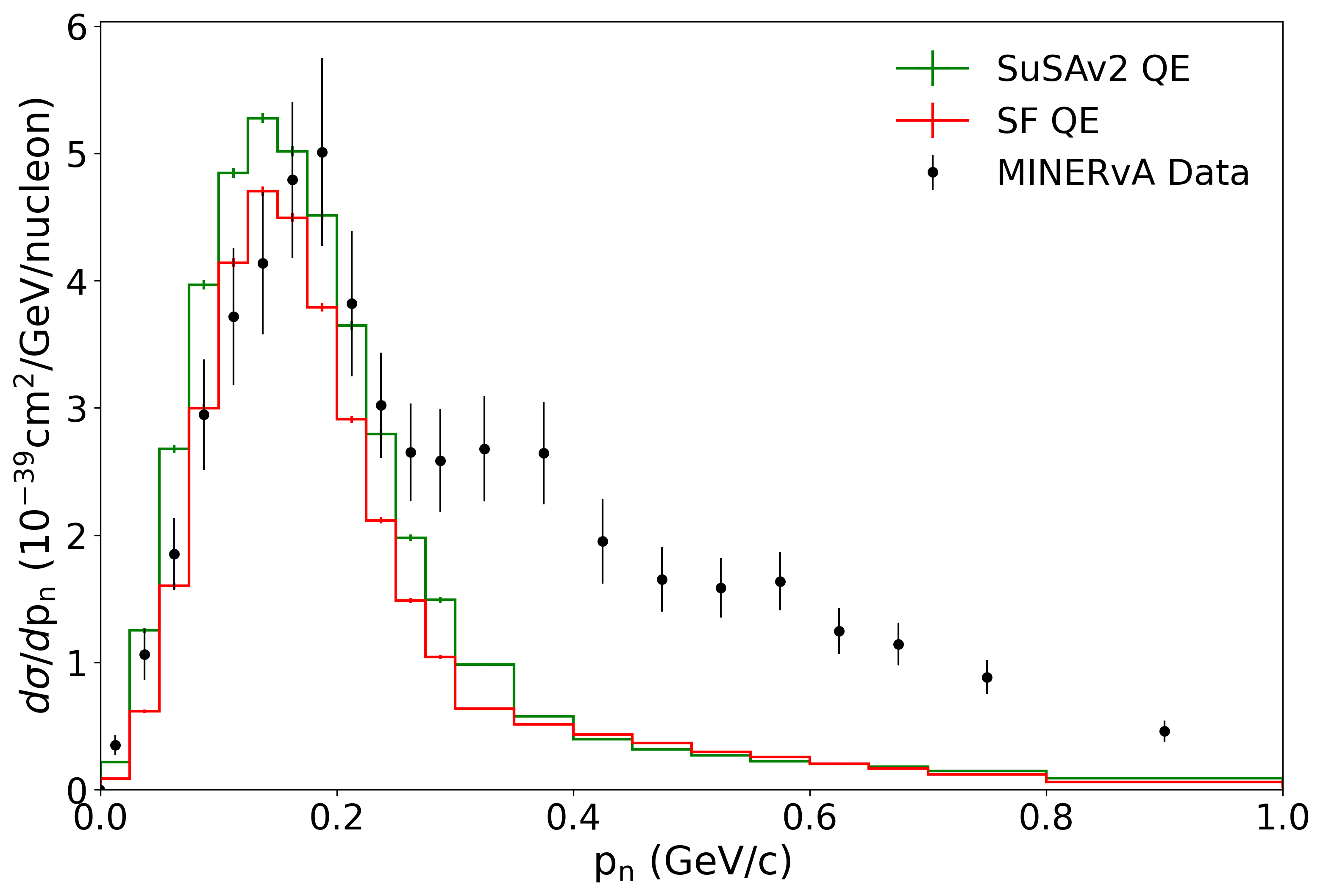}
    \caption{MINERvA differential cross section in $p_{n}$ (initial neutron momentum) with data in (black points) compared to SF (red), and SuSAv2 (green). Data from~\cite{MINERvA:2018hba}.}
    \label{fig:pn}
\end{figure}
The predicted $p_{n}$ distribution from the SF matches the data very well in width and peak position and is slightly narrower than the SuSAv2 prediction, which reflects the broader initial nucleon momentum distribution of the LFG used by SuSAv2 in GENIE as seen in Fig.~\ref{fig:momentum_dist}. 
\vskip 0.12in
The same analysis measured the leading proton scattering angle spectrum, which is sensitive to FSI but also to the way in which the final state nucleon phase space is sampled~\cite{Nikolakopoulos:2023pdw}. As the SuSAv2 implementation is inclusive there is no guarantee that the final state nucleon kinematics will be correctly generated, as opposed to the fully exclusive nature of the SF implementation. 
\begin{figure}[h]
    \centering
    \includegraphics[width=\linewidth]{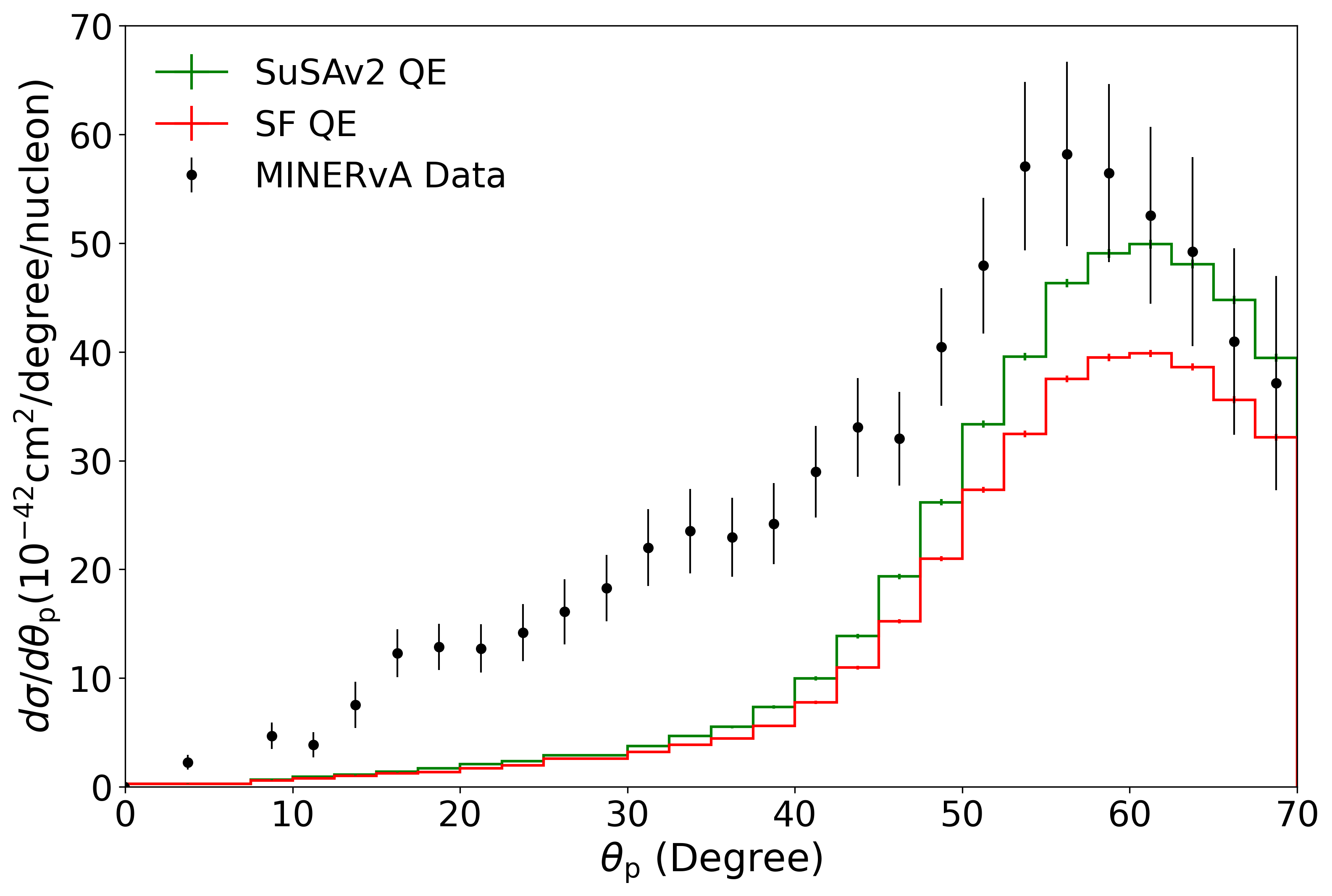}
    \caption{MINERvA differential cross section in $\theta_{p}$ (proton scattering angle) with data in (black points) compared to SF (red), and SuSAv2 (green). Data from~\cite{MINERvA:2018hba}.}
    \label{fig:p_theta}
\end{figure}
Figure~\ref{fig:p_theta} shows the proton scattering angle spectrum, with the SuSAv2 QE prediction being significantly larger and slightly broader at the QE peak than the SF prediction. As 2p2h and other inelastic channels will contribute over the entire range of proton scattering angles, the SuSAv2 prediction leads to an over-estimation of the cross section. 
\vskip 0.12in
The final MINERvA data set for comparison is the triple differential CCQE-like measurement in the medium energy, with an average neutrino energy of $6$ GeV~\cite{MINERvA:2022mnw}. In this analysis, data is binned in muon longitudinal and transverse momentum as well as $E_{\rm{avail}}$ defined by
\begin{equation}\label{eq:Eavail}
    E_{\rm{avail}} = \sum T_{\rm{proton}} + \sum T_{\pi^{\pm}} + \sum E_{\rm{particle}}\, .
\end{equation}
In the above, $T_{\rm{proton}}$ is the proton kinetic energy, $T_{\pi^{\pm}}$ is the charged pion kinetic energy, and $E_{\rm{particle}}$ is the total energy of any other final state particle except neutrons~\cite{MINERvA:2022mnw}. This kinematic variable when summed with the lepton energy is used as an estimator for the true neutrino energy by experiments like NOvA and MicroBooNE. In this measurement the signal is $0\pi$ events, so $E_{\rm{avail}}$ is just the sum of the kinetic energies of all detected protons. 
\vskip 0.12in
Figure~\ref{fig:triple} shows this triple differential cross section for $1.5\,\rm{GeV} < p_{||} < 3.5\,\rm{GeV}$ with QE predictions from the SF, SuSAv2, and G2018 models. As this sample contains high energy neutrinos, there is again the expectation that each simulation's prediction should undershoot the data, which we see from each of the three models. It is interesting to note that even though each of the three included models have vastly different theoretical underpinnings, that each lead to a similar prediction in the QE region. 
\vskip 0.12in
Our last data comparison is with T2K data on oxygen. An oxygen Spectral Function computed using CBF theory has also been provided in GENIE, enabling the SF model to be validated against multiple nuclear targets~\cite{Benhar:1994hw}. We compare against double differential cross sections in muon momentum and cosine of the scattering angle on oxygen from $CC0\pi$ events from T2K~\cite{T2K:2020jav}. The lower beam energy of T2K, which peaks at around $600$ MeV means that inelastic contributions from resonance production and DIS are smaller than in other neutrino experiments. 
\begin{figure*}[t]
    \centering
    \includegraphics[width=\linewidth]{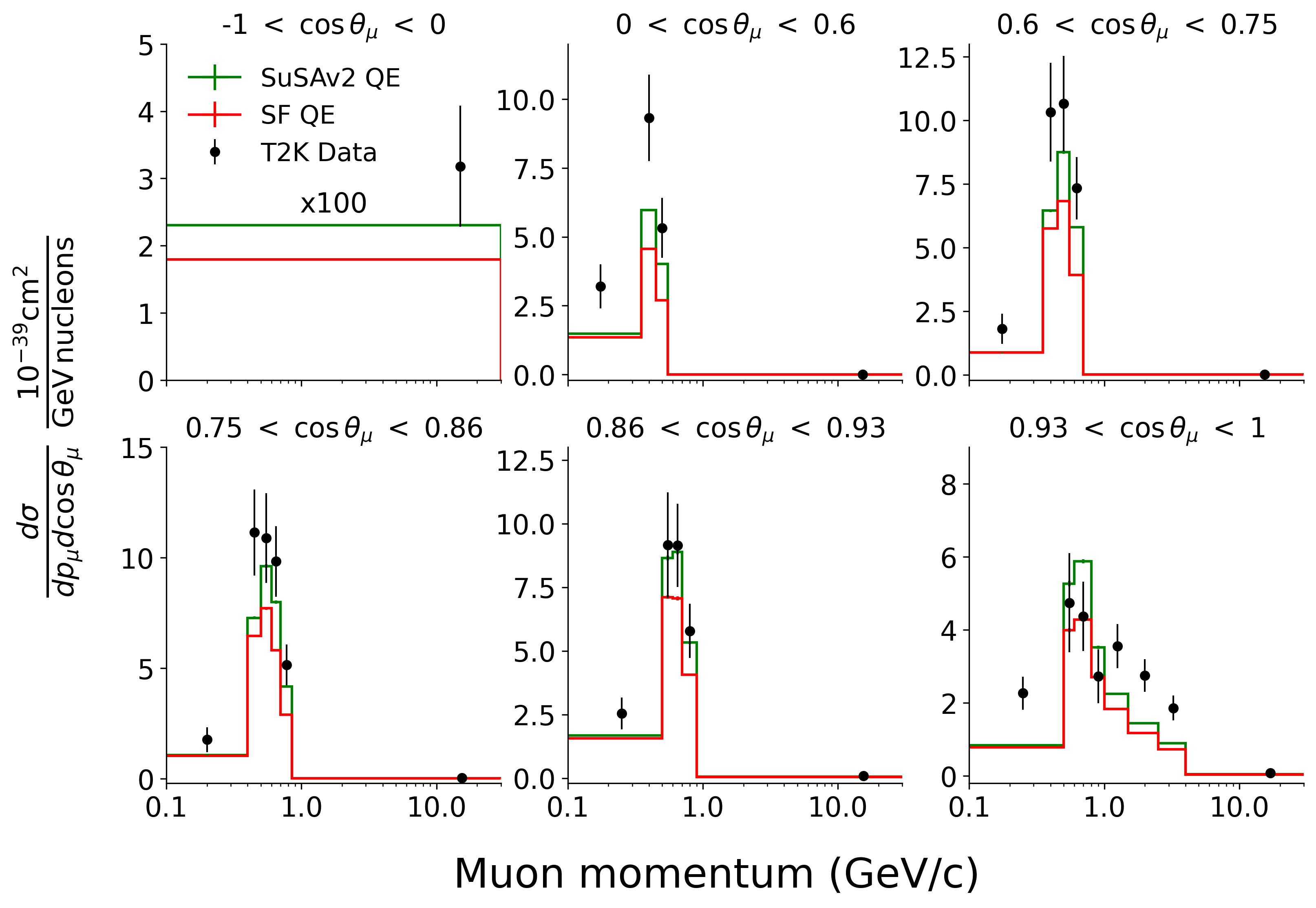}
    \caption{T2K double differential cross section in muon momentum and cosine of the muon scattering angle per nucleon on $^{16}$O. Results are shown with data in black points, the GENIE SuSAv2 QE only prediction (green) and GENIE SF QE only prediction (red). Note the logarithmic x axis. }
    \label{fig:t2k_oxygen}
\end{figure*}
In Figure~\ref{fig:t2k_oxygen} we compare predictions from the SF and SuSAv2 models in 6 different bins of $\cos\theta_{\mu}$. SF predictions are consistently below the data, as to be expected as 2p2h interactions are still expected to be significant at these kinematics as well as small contributions from resonance production. This is to be compared to the SuSAv2 QE predictions which are already close to the data and even overshoot it at forward muon angles.

\section{Discussion}
The growing quantity, quality, and dimensionality of charged lepton and neutrino-nucleus scattering data present increasingly strong constraints on event generator predictions.
To meet the precision simulation needs of future experiments, an efficient pathway for the implementation of more realistic, theory driven models which start from a microscopic picture of the nucleus will be invaluable. We have highlighted some practical difficulties in including such new models in neutrino event generators like GENIE, and we have created an interface for Fortran-based QE cross-section calculations as a first step to 
overcome these difficulties. 

We have also discussed some of the limitations of the available models in GENIE, focusing on the highly approximate representations of the nuclear ground state currently available. We have shown how the Spectral Function provides a more complete picture of the nucleus with the correct relationship between nucleon momentum and removal energy, as well as naturally including correlations between nucleons. This more complex model for the nuclear ground state leads to marked differences in exclusive cross-section predictions as can be seen in both electron and neutrino scattering.

Finally, the inclusion of the Spectral Function model within GENIE allows for multiple avenues for continuing improvement. The first is the ability of this model, and the code as implemented, to predict electron and neutrino scattering cross sections simultaneously. This will allow information gathered from precision charged lepton scattering experiments to be more effectively used to refine neutrino scattering predictions. While this work is limited to the quasielastic region, it is important to mention that the SF formalism has been generalized to include two-body current and pion-production mechanisms~\cite{Rocco:2019gfb}. However, the process of extending the interface to encompass these additional contributions is not straightforward and would necessitate further advancements beyond the present scope.

In contrast to similar previous efforts based on precomputed hadron tensors, the interface we have devised provides greater ability to manipulate input parameters and study their impact on the simulation predictions. In particular, our interface allows for estimation of theoretical uncertainties through direct variations of the adopted nucleon form factors and the use of multiple Spectral Function tables calculated using different nuclear model assumptions~\cite{Simons:2022ltq,universe9080367}. 


\section{Acknowledgments}
This manuscript has been authored by Fermi Research Alliance, LLC under Contract No. DE-AC02-07CH11359 with the U.S. Department of Energy, Office of
Science, Office of High Energy Physics and Fermilab LDRD awards (N.S).

\bibliography{SF}

\end{document}